\documentclass[%
superscriptaddress, twocolumn,
 amsmath,amssymb,
 aps,pra
]{revtex4-1}

\usepackage{braket}
\usepackage{graphicx,bm,libertine}
\usepackage{comment}
\usepackage[dvipsnames]{xcolor}
\usepackage{subfigure}
\usepackage[utf8]{inputenc}
\usepackage{placeins}
\usepackage[colorlinks=true,linkcolor=blue,urlcolor=blue,citecolor=blue,pdfusetitle]{hyperref}

\newcommand{\one}{\mbox{$1 \hspace{-1.0mm}  {\bf l}$}}

\begin{document}

\title{Macroscopic quantumness of optically conditioned mechanical systems}

\author{Hannah McAleese}

\author{Mauro Paternostro}

\affiliation{Centre for Theoretical Atomic, Molecular and Optical Physics, School of Mathematics and Physics, Queen's University, Belfast BT7 1NN, United Kingdom\\
}%

\date{\today}

\begin{abstract}
We address the macroscopic quantumness of the state of mechanical systems subjected to conditional protocols devised for state engineering in cavity optomechanics. We use a measure of macroscopicity based on phase-space methods. We cover the domain from weak to strong optomechanical coupling, illustrating how measurements performed over the cavity field that drives the dynamics of a mechanical system are able to steer the latter towards large quantum coherent states. The effect of losses is evaluated for the case of an open cavity and analyzed in terms of the features of the Wigner functions of the state of the mechanical system. We also address the case of engineered phonon-subtracted mechanical systems, in full open-system configuration, demonstrating the existence of optimal working points for the sake of mesoscopic quantumness. Our study is relevant for and applicable to a broad range of settings, from clamped to levitated mechanical systems. 
\end{abstract}

\maketitle

Ever since Schrodinger’s famous thought experiment \cite{Schrodinger1935}, the concept of quantum superpositions in macroscopic systems has led many, both inside and outside the scientific community, to reconsider our understanding of physical reality. There remain numerous open questions about how the size of a system is related to its ability to exhibit quantum behavior. Macroscopic quantum states can be used to bring us closer to answering these questions and to understanding more clearly the quantum-classical boundary \cite{Zurek2003}. In addition, states of this kind are valuable resources for quantum computation and quantum metrology \cite{FrowisReview}.

The identification of successful strategies for the generation of such macroscopic quantum states is an active area of investigation. Promising avenues in this direction have been explored in the field of cavity optomechanics~\cite{AspelmeyerReview,CavityMirror, Bose1999, Marshall2003, Kleckner2008, Li2013, Carlisle2015}, owing to the possibility offered by such a platform to transfer the quantumness of the state of a light field driving the motion of a mechanical system to the state of the latter. The overarching scope of such endeavors is to take mesoscopic quantum states one step closer to experimental realization. 

However, a currently open question is whether it is possible to define a comprehensive figure of merit able to {\it quantify} the degree of macroscopic quantumness of a given state. A range of measures of macroscopic quantumness have been proposed \cite{Leggett80, Leggett2002, Dur2002, Shimizu2002, Bjork2004, Cavalcanti2006, Cavalcanti2008, Marquardt2008, Macro, Park2016, Frowis2012, Nimmrichter2013, Sekatski2014, Sekatski2017, Laghaout2015, Yadin2015, Kwon2017, Schrinski2019}, based on different definitions and means of quantifying this concept. Such measures vary in regard to the physical systems to which they can be applied, the states for which they are relevant, and the features that characterize a given macroscopically quantum state. 

In this paper, we focus on an optomechanical setting for the engineering of macroscopic quantum states and make use of the measure proposed in Ref.~\cite{Macro}. Such a phase-space based quantifier is instrumental to our purposes in light of the possibility to apply it to a broad range of both pure and mixed states of a mechanical system. 

With this choice of system and quantifier of macroscopic quantumness in mind, we investigate a scheme for the preparation of mesoscopic quantum states of a mechanical system that is assisted by measurements performed on the optical field that drives the mechanical motion. We study the impact that measurements have on the resulting degree of mechanical macroscopicity, analyzing the influences of the specific measurement settings on the resultant mechanical state. We then use the results of such analysis to find effective measurement strategies to engineer mechanical states with enhanced degrees of macroscopicity.

The remainder of this paper is organized as follows. In Sec.~\ref{sec:optosys} we describe the systems that we consider. We then give in Sec.~\ref{sec:measure} an introduction to the measure of macroscopic quantumness used to characterize the performance of the schemes that we analyze. In Sec.~\ref{sec:hdetMirror} we illustrate the effect of conditional strategies on the state of the mechanical system and quantify the degree of macroscopic quantumness of the latter when performing either homodyne or heterodyne optical measurements, seeking for the optimal measurement strategy for the sake of macroscopicity. Sec.~\ref{sec:noiselin} is dedicated to the analysis of the effect of cavity damping on the degree of macroscopicity achieved through such a scheme. Sec.~\ref{sec:subtraction} illustrates an alternative approach to the establishment of mechanical macroscopic quantumness based on photon subtraction. Finally, in Sec.~\ref{sec:conc} we provide our concluding remarks, while part of our formal analysis is reported in a technical Appendix. 

\section{Optomechanical Systems}
\label{sec:optosys}

We consider a standard optomechanical system, which might consist of a single-sided cavity endowed with a clamped mechanical nanostructure as well as a standard cavity accommodating a levitated nanoparticle. Regardless of the specific configuration of the system, the Hamiltonian can be cast in the form~\cite{CavityMirror}
\begin{equation} \label{HamLinear}
    H = \hbar \omega_o a^\dagger a + \hbar \omega_m b^\dagger b - \hbar g_l a^\dagger a (b + b^\dagger),
\end{equation}
where $\omega_o$ ($\omega_m$) is the frequency of the cavity field (mechanical mode) whose annihilation and creation operators are $a$ and $a^\dagger$ ($b$ and $b^\dagger$). The rate $g_l= ({\omega_o}/{L}) \sqrt{{\hbar}/{(2 m \omega_m)}}$ quantifies the vacuum optomechanical coupling strength. Here $L$ is the length of the cavity and $m$ is the effective mass of the mechanical oscillator. 

The time evolution operator corresponding to the Hamiltonian in Eq.~(\ref{HamLinear}) can be rearranged as
\begin{equation}
\label{tevol}
    U(\tau) = e^{-i b^\dagger b \tau}e^{-i r a^\dagger a \tau} e^{i k^2 (a^\dagger a)^2 [\tau - \sin(\tau)]} D_m(k\eta a^\dag a)
    ,
\end{equation}
where the relevant parameters have been rescaled and cast in a dimensionless form as $r = \omega_o/\omega_m$, $k = g/\omega_m$ and $\tau = \omega_m t$ with $t$ the evolution of the system. The term $\eta$ relates to the dimensionless evolution time as $\eta = 1-e^{-i\tau}$. We have introduced the displacement operator $D_m(x)=\exp[x b^\dag-x^* b]$, which displaces the state of the mechanical mode in the phase space. Notice that the amplitude of the displacement performed in Eq.~\eqref{tevol} depends on the number of excitations in the state of the cavity field through the number operator $a^\dag a$.

As in Ref.~\cite{CavityMirror}, we take the initial state of the cavity field and mirror to be coherent states $\ket{\alpha}_c$ and $\ket{\beta}_m$. This can be achieved by feeding the cavity with an input pump at frequency $\omega_p$. We call $\Delta=\omega_o-\omega_p$ the detuning between pump and cavity field: for a resonant coupling between cavity field and external pump (i.e. for $\Delta=0$), the state of the former is asymptotically driven to a coherent state of amplitude determined by the intensity of the latter. As for the mechanical system, coherent states can be engineered with large purity using pulsed-driving schemes as in Refs.~\cite{Vanner,Rossi}. For this initial preparation, the time-evolved state of the system is
\begin{equation} \label{timeEvState}
\ket{\psi(\tau)} = e^{-|\alpha|^2/2} \sum_{n=0}^{\infty} c_n (\alpha, \tau) \ket{n}_c \ket{\phi_n (\tau)}_m,
\end{equation}
where $\{\ket{n}_c\}$ are number states of the cavity field, $\ket{\phi_n (\tau)}_m$ is a coherent state of the mechanical system with amplitude $\phi_n (\tau) = k \eta n + \beta e^{-i \tau}$ and we have introduced the coefficients 
\begin{equation}
c_n (\alpha, \tau) = \frac{(\alpha e^{\varphi })^n}{\sqrt{n!}} e^{i k^2 n^2 [\tau-\sin(\tau)] - ir \tau n}
\end{equation}
with $\varphi = k(\eta \beta^* e^{i \tau} - \eta^* \beta e^{-i \tau})/2$. As the displacement of the mechanical mode depends on the number of photons in the cavity, Eq.~\eqref{timeEvState} describes, in general, an inseparable state.

\section{Measure of Macroscopic Quantumness}
\label{sec:measure}

We now briefly review the measure of macroscopic quantumness that has been used in our quantitative assessment. Following Ref.~\cite{Macro}, we consider the quantity
\begin{equation} \label{macroEq}
\mathcal{I} (\rho) = \frac{1}{2 \pi} \int (|\gamma|^2 - 1) |\chi(\gamma)|^2 d^2 \gamma, 
\end{equation}
where $\rho$ is the density matrix of the system at hand, $\chi(\gamma) = \mathrm{Tr} [\rho D(\gamma)]$ is the Weyl characteristic function of state $\rho$~\cite{Alessandro} and $\gamma\in\mathbb{C}$ is a phase-space variable. The first term in the integral (depending on $|\gamma|^2$) quantifies the magnitude of phase-space interference fringes that result from the possible quantum coherence inherent in the state of the system. The square modulus of the characteristic function, on the other hand, measures the frequency of such fringes and thus the size of the superposition state at hand~\cite{JeongReview}. Eq.~\eqref{macroEq} thus allows us to quantify the degree of quantum coherence and size of the system simultaneously. Moreover, at variance with other quantifiers that have been recently proposed to determine the degree of macroscopicity of a quantum state, such quantity can be evaluated without the need for difficult optimization or special decompositions of the state $\rho$ under scrutiny. As ${\cal I}(\rho)$ is based on the phase-space features of a state, it is well suited to address the macroscopic quantumness of bosonic systems and, in general, all those systems whose phase-space behavior is readily available. The maximum value that $\mathcal{I}(\rho)$ can take is the average number of particles or excitations in the state of the system $N_\rho$. Such upper bound is achievable only by pure states, mixed ones giving ${\cal I}(\rho)< N_\rho$, strictly~\cite{JeongReview}. As we are interested in characterizing the macroscopic quantumness of the mechanical system, in the remainder of this work we shall take $N_\rho=\langle{b^\dagger b}\rangle$.

\section{Conditioning the state of the mechanical system} 
\label{sec:hdetMirror}

Eq.~\eqref{timeEvState} shows that the reduced state of the mechanical oscillator consists of a convex sum of coherent states, which will then exhibit no quantum feature. However, the entanglement that is in general shared by cavity field and mechanical system allows to condition the state of the latter by means of controlled quantum operations on the former. In particular, the performance of projective measurements onto the state of the cavity field provides a powerful tool for the achievement of potentially interesting states of the mechanical system, when gauged against their degree of non-classicality~\cite{CavityMirror,Li2013}. 

\begin{figure}[t]
    \centering
{\bf (a)}
    \includegraphics[width=\columnwidth]{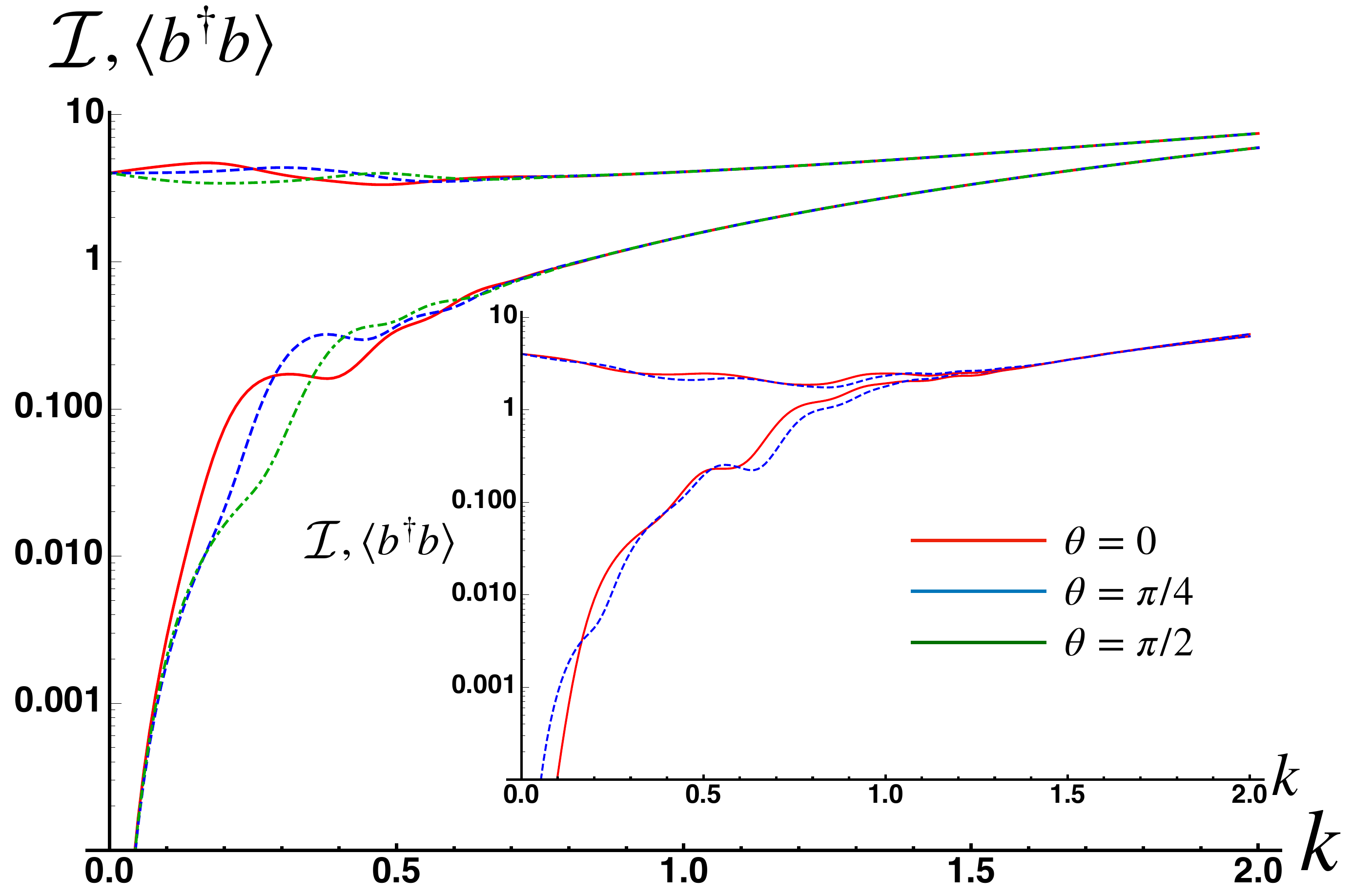}\\
    {\bf (b)}
     \includegraphics[width=\columnwidth]{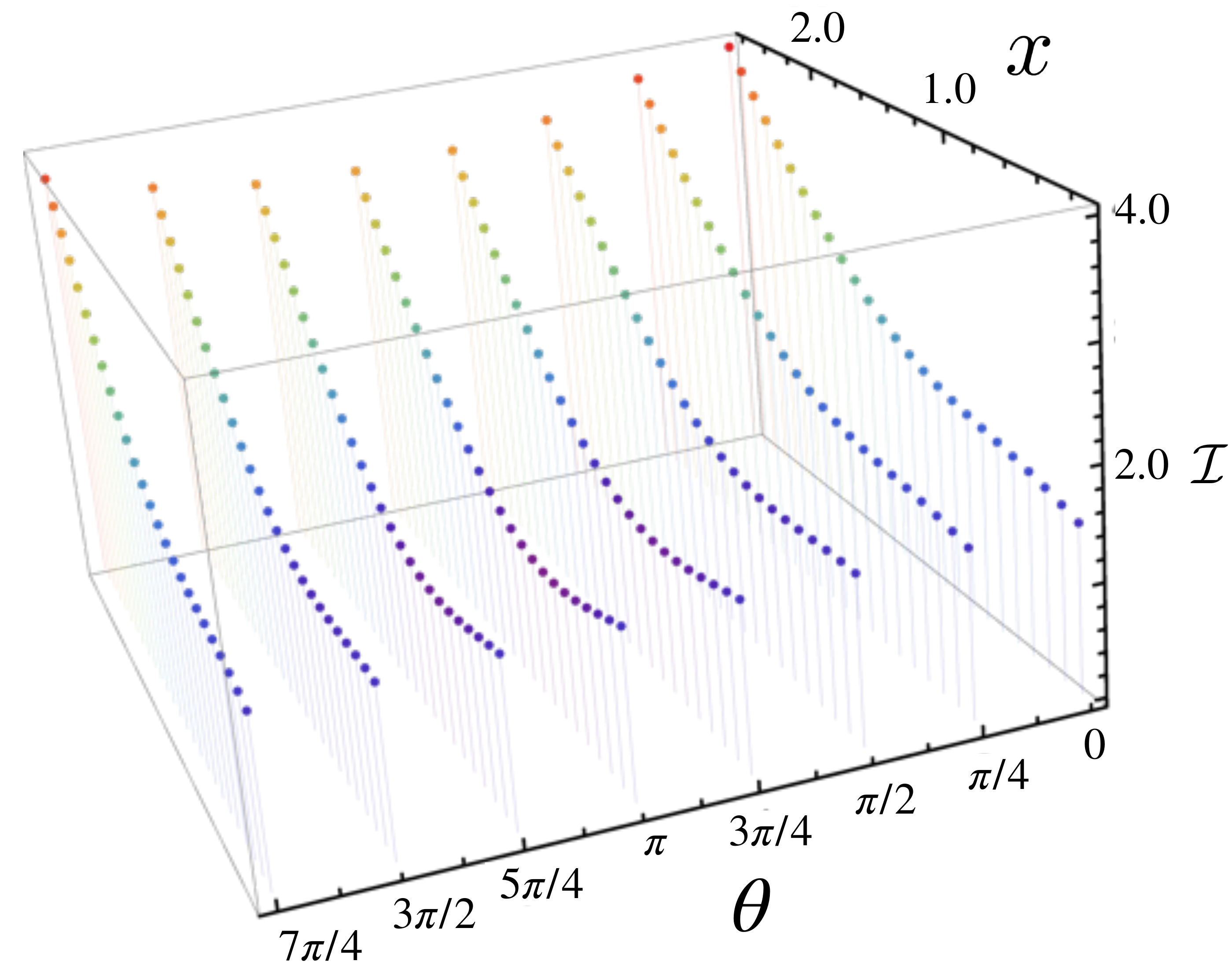} 
    \caption{{\bf (a)}: We plot the measure of macroscopicity ${\cal I}$ and the mean number of excitations $\langle b^\dag b\rangle$ against the effective coupling strength $k$ for homodyne detection with $\theta=0$ (solid red curves), $\theta=\pi/4$ (dashed blue curves) and $\theta=\pi/2$ (dot-dashed dark green curves). We have used $\alpha=0.8$, $\beta=2$, $x=0$ and $\tau=\pi$ in our calculations. Inset: same analysis but for $x=1$ and limited to $\theta=0,\pi/4$ only. {\bf (b)}: We study the behavior of ${\cal I}$ against $x$ and $\theta$ for $k=1$. All other parameters as in panel {\bf (a)}. }
    \label{fig:hdyneTheta}
\end{figure}

 In order to generate a quantum state of the mirror, we thus need to subject the field to a measurement that results in the superposition of the coherent states that appear in Eq.~\eqref{timeEvState}. This rules out photon-counting as a suitable strategy to implement in light of the form of the joint optomechanical states. We thus resort to general-dyne measurements encompassing Gaussian measurements, which can be modeled as von Neumann projections of the cavity field onto suitable Gaussian states~\cite{Alessandro}. 
 
 In particular, ideal homodyning along arbitrary directions in phase space implies the projection onto the eigenstates $\ket{x(\theta)}$ of the quadrature $x(\theta)=a e^{i\theta}+h.c.$, i.e.
 \begin{equation}
 \ket{x(\theta)}=e^{-x^2/2}\sum^\infty_{n=0}\frac{e^{i\theta n}}{\sqrt{2^n n!}\pi^{1/4}}H_n(x)\ket{n}
 \end{equation}
 with $H_n(x)$ the Hermite polynomials of order $n$ and variable $x\in[-\infty,\infty]$ and $\theta\in[0,2\pi]$ a phase-space angle. As for ideal heterodyne, they are {\it de facto} modelled as projections onto coherent states $\ket{\sigma}$ with $\sigma\in\mathbb{C}$. In what follows, we will use the notation $M=\text{hom}$ ($M=\text{het}$) to indicate a homodyne (heterodyne) measurement.
 
 Regardless of the specific form of the measurement being implemented, the conditional state of the mechanical mode (after normalization) can be cast in the general form
 \begin{equation}
 \label{general}
 \ket{\psi'}_m={\cal N}_Me^{-|\alpha|^2/2}\sum^\infty_{n=0}d_n(\alpha,\tau)\ket{\phi_n(\tau)}_m
 \end{equation}
 with ${\cal N}_M$ a measurement-dependent normalization constant whose form is immaterial, and $d_n(\alpha,\tau)=c_n(\alpha,\tau) f_M(n)$, where
 \begin{equation}
 \label{fM}
 f_M(n)=
 \begin{cases}
 \frac{1}{\sqrt{n!}} e^{-|\sigma|^2/2}\sigma^{*n}~~~~~~~~~~~~~~~\text{for $M=$ het},\\
 \frac{1}{\sqrt{2^n n!}\sqrt[4]{\pi}} e^{-x^2/2-i\theta n} H_n(x)~\text{for $M=$ hom}.
 \end{cases}
 \end{equation}
 
 In turn, the characteristic function of such conditional states can be written in a general fashion as
 \begin{equation}
 \chi(\gamma)={\cal N}^2_M\sum^\infty_{n,l=0}e^{-|\alpha|^2}{\cal M}_{n,l}(\alpha,\tau){\cal F}_{n,l}(\gamma,\tau)
 \end{equation}
 with ${\cal M}_{n,l}(\alpha,\tau)=d_n(\alpha,\tau)d^*_l(\alpha,\tau)$, which is fully determined by the choice of measurement being performed, and ${\cal F}_{n,l}(\alpha,\tau)$ the {\it universal} (i.e. measurement independent) function
 \begin{equation}
 \begin{aligned}
 {\cal F}_{n,l}(\gamma,\tau)&={}_m\langle\phi_l(\tau)|D(\gamma)|\phi_n(\tau)\rangle_m\\
 &=e^{-(|\phi_l(\tau)|^2+|\gamma+\phi_n(\tau)|^2)/2+\gamma\phi^*_n(\tau)-\gamma^*\phi_l(\tau)}.
\end{aligned}
 \end{equation}
 As this is the only $\gamma$-dependent part of the characteristic function, it is the only term contributing to the phase-space integral inherent in the definition of the measure of macroscopic quantumness. We shall thus consider the integral
 \begin{equation}
 \begin{aligned}
& \int(|\gamma|^2{-}1){\cal F}_{n,l}(\gamma,\tau){\cal F}^*_{p,q}(\gamma,\tau)\,d^2\gamma=\pi  [\phi_l(\tau)-\phi_p(\tau)]^*\\
 &\times [\phi_q(\tau){-}\phi_n(\tau)] e^{ \phi^*_l(\tau) \phi_q(\tau)+\phi_n(\tau) \phi^*_p(\tau)-\frac{1}{2} \sum_{j=l,n,p}|\phi_j(\tau)|^2},
\end{aligned}
 \end{equation}
 which would enter Eq.~\eqref{macroEq} to account for the contribution arising from the superposition of coherent states in Eq.~\eqref{general}.
 
 \subsection{Effect of Homodyne Detection} \label{homolinear}

 With this result at hand, we can now evaluate measure ${\cal I}$ for different choices of the measurement strategy. Motivated by Ref.~\cite{CavityMirror}, we shall start considering the case of ideal and general homodyne measurements of the field, aiming at identifying the role played by the phase-space angle $\theta$, the {\it effective interaction strength} $k$ and the value $x(\theta)$ of the homodyne signal. 

In Fig.~\ref{fig:hdyneTheta} {\bf (a)} we address the effect that a growing value of $k$ has on the measure of macroscopicity for various choices of the phase-space angle $\theta$ and for a homodyne measurement with postselection of the $x=0$ outcome. Only at small values of $k$ is the latter parameter relevant in the determination of the degree of macroscopicity. However, there is no clear hierarchy among the values of ${\cal I}$ associated with a growing $\theta$ as the curves corresponding to different choices of the phase space angle are very intertwined and cross each other at multiple points. As $k$ grows, such curves tend towards a monotonic behavior with virtually no dependence on the specific choice of $\theta$ made for the measurement setting. The region of large values of the effective interaction strength also corresponds to a shrinking gap with the mean number of excitations in the mechanical mode $\langle b^\dag b\rangle$, thus providing evidence of a stronger macroscopically quantum behavior of the conditional state of the mechanical mode. The relation among the trends followed for the various choices of values of $\theta$ considered in our study is found also in the behavior of the mean number of excitations, thus showing that ${\cal I}$ is able to capture -- in a faithful manner -- intrinsic features of the states we have been considering. These considerations are found to hold also for $x\neq0$, as shown in the example provided by the inset of Fig.~\ref{fig:hdyneTheta} {\bf (a)}, which is for $x=1$. While non-zero values of $x$ deliver the same qualitative results as for $x=0$, the value of $k$ starting from which any dependence on $\theta$ is apparent is larger for $x\neq0$ than for projections onto the origin of the phase space. Moreover, the value of $k$ at which ${\cal I}\simeq\langle b^\dag b\rangle$ appears to bear a dependence on the choice of $x$.

\begin{figure}[t]
    \centering
    {\bf (a)}
\includegraphics[width=\columnwidth]{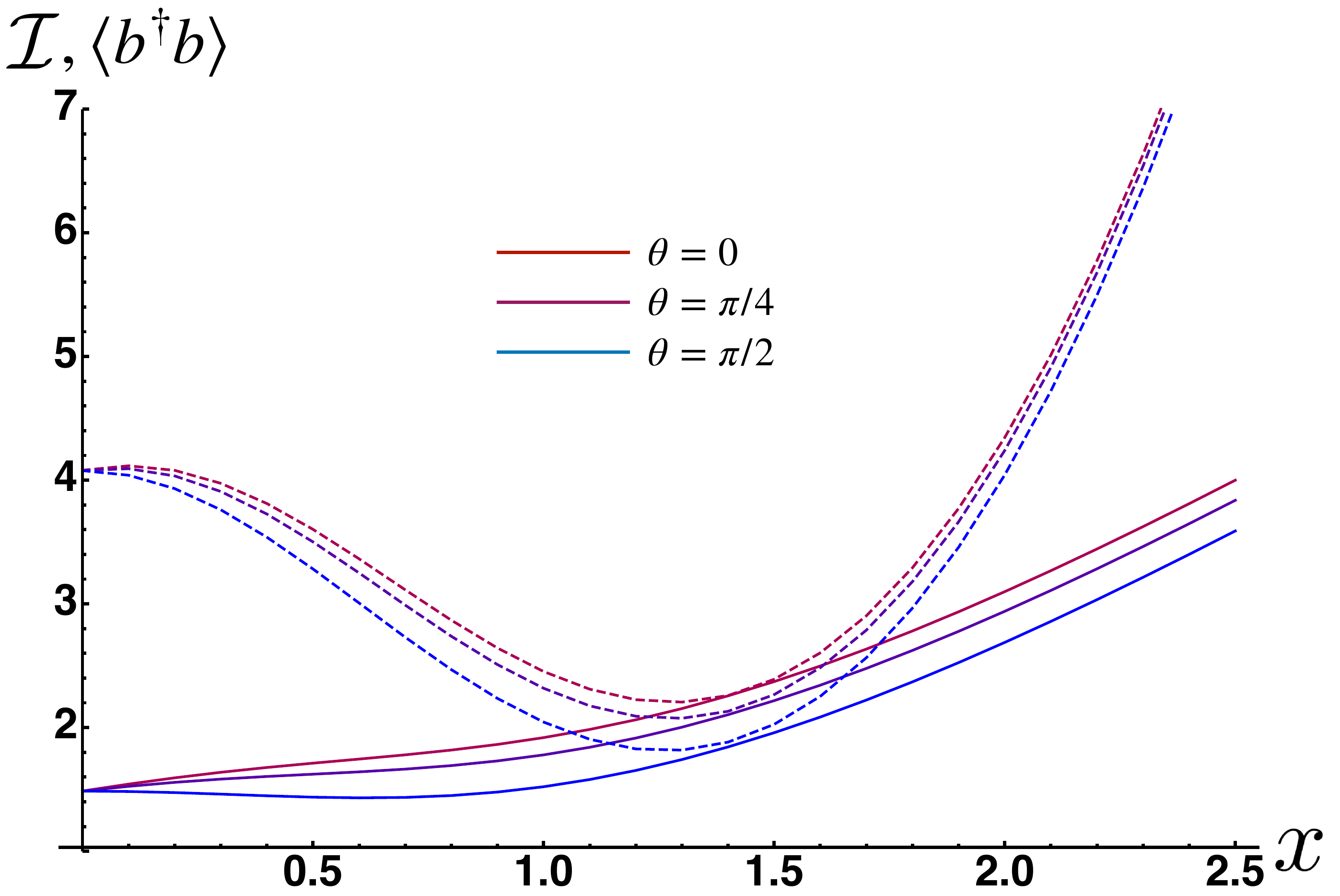}\\
{\bf (b)}
\includegraphics[width=\columnwidth]{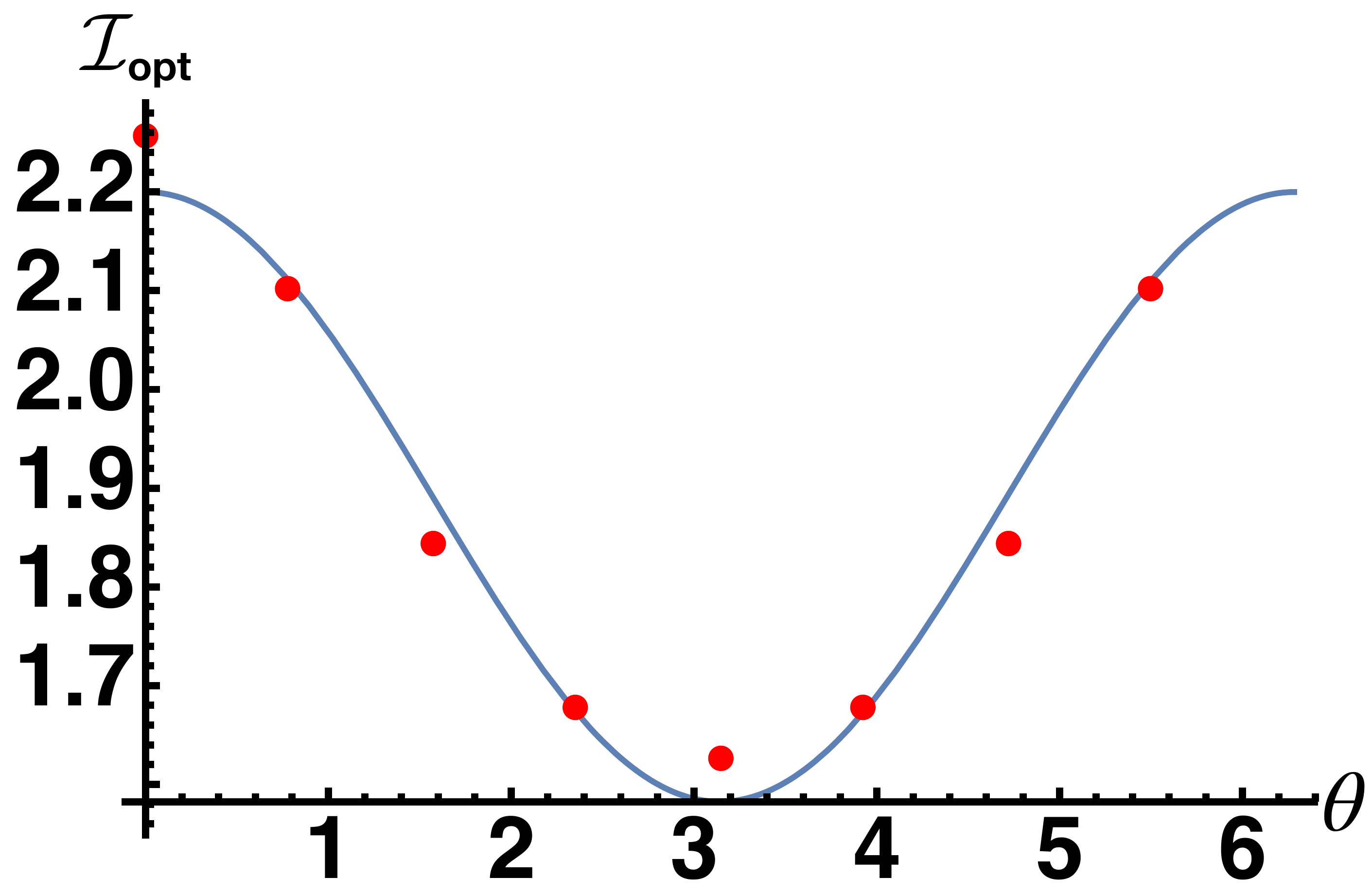}
    \caption{{\bf (a)} We study the behavior of ${\cal I}$ (solid lines) and $\langle b^\dag b\rangle$ (dashed lines) against $x$ for a homodyne measurement with three values of $\theta$ as per the legend.  {\bf (b)} Analysis of ${\cal I}_\text{opt}$ against the phase-space angle $\theta$ for $\alpha=0.8$, $\beta=2$, $x=1.42701$ and $k=1$. The solid line is a sinusoidal best fit trend of the form $a+c \sin(\theta+b)$ with $a=1.891$, $b=\pi/2$ and $c=0.309$.} 
    \label{fig:hdyneMean}
\end{figure}

In order to investigate further the influence that various choices of $\theta$ and $x$ have on the degree of macroscopicity, we now fix the value of $k$. As a compromise between the size of our computation and the degree of macroscopicity that we aim at achieving, in what follows and unless otherwise specified, we consider $k=1$, which corresponds to a sizeable macroscopic superposition and a working point where the dependence on $\theta$ is less complex than for lower values of such parameter. Fig.~\ref{fig:hdyneTheta} {\bf (b)} shows that, as $x$ grows, a non-zero value of $\theta$ is detrimental to the degree of macroscopicity being achieved through homodyne measurements, although the variations in ${\cal I}$ are  small. This suggests that, regardless of the value of the homodyne signal being post-selected, the best strategy is to project onto eigenstates of the position quadrature. This is further illustrated in Fig.~\ref{fig:hdyneMean} {\bf (a)}, where we study the relation between ${\cal I}$ and the mean number of bosonic excitations against $x$ and for a few values of $\theta$, showing the general decreasing trend of the degree of macroscopicity with a growing phase-space angle and the existence of a $\theta$-independent value of the homodyne signal $\bar{x}=1.42701$ at which ${\cal I}\simeq\langle b^\dag b\rangle$. We shall call ${\cal I}_\text{opt}$ the degree of macroscopicity at this value of $\bar{x}$. Fig.~\ref{fig:hdyneMean} {\bf (b)} shows that ${\cal I}_\text{opt}$ follows an approximatively sinusoidal trend. 

\begin{figure*}[t!]
    \centering
    {\bf (a)}\hskip5.5cm{\bf (b)}\hskip5.5cm{\bf (c)}
    \includegraphics[width=2\columnwidth]{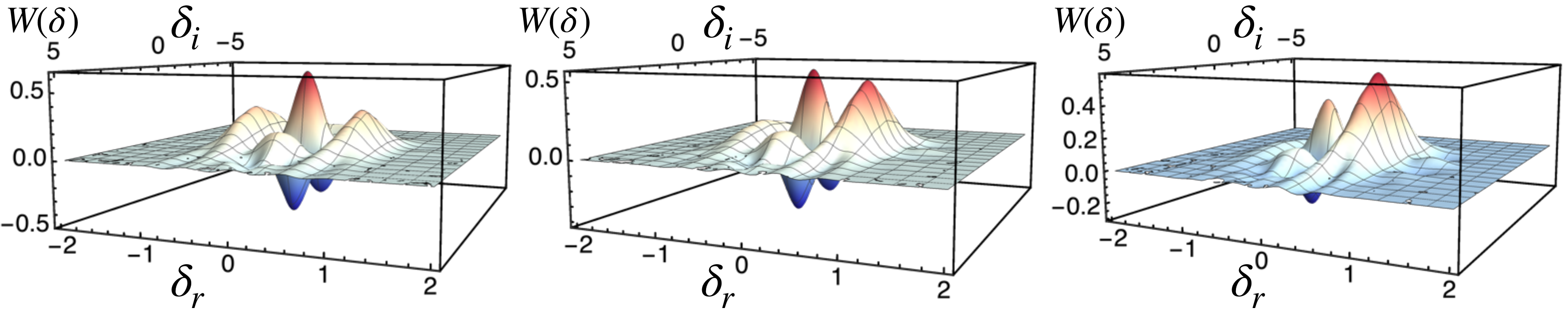}
    \caption{Wigner function of $\rho_B$ when $x=1.42701$. It looks like a cat state, with interference between two positive states. I have taken $\alpha=0.8$, $\beta=2$, $k=1$ and $\tau=\pi$.}
    \label{fig:hdyneWigner}
\end{figure*}

Having identified a potentially optimal working point for the case of homodyne measurements, we now aim at characterizing the state of the mechanical mode achieved correspondingly. To this goal, we focus on the Wigner function of the mechanical mode at $x=\bar{x}$, $\theta=0$ and the parameters used in Fig.~\ref{fig:hdyneTheta} (which are motivated by the analysis reported in Ref.~\cite{CavityMirror}). The Wigner function of a single harmonic oscillator is defined as 
\begin{equation}
    W(\delta) = \frac{1}{\pi^2} \int e^{\gamma^* \delta - \gamma \delta^*} \chi(\gamma) d^2 \gamma
\end{equation}
with $\delta\in\mathbb{C}$. Fig.~\ref{fig:hdyneWigner} {\bf (a)} shows significant quantum coherence in the state of the mechanical mode, as witnessed by the large interference fringes (taking both positive and negative values) between two symmetrically placed Gaussian distributions. These features, which are reminiscent of those of a Schr\"odinger cat state having the form of a quantum superposition of coherent states -- a similarity that we will address quantitatively later on in this paper -- account for the macroscopically quantum character witnessed by ${\cal I}$. The analysis of the Wigner function allows us to gather an intuition of the depleted degree of macroscopic quantumness resulting from non-zero values of $\theta$. In Fig.~\ref{fig:hdyneWigner} {\bf (b)} and {\bf (c)} we report $W(\delta)$ for $\theta=\pi/4$ and $\pi/2$ respectively, showing that the symmetry that is characteristic of the case with $\theta=0$ is lost. This suggests that the effect of $\theta$ is that of biasing the quantum superposition towards one component, thus strongly depleting the other one and justifying intuitively the decrease of ${\cal I}$ observed in Fig.~\ref{fig:hdyneMean} {\bf (a)}.

In order to address the similarity of the mechanical mode state to a quantum superposition of displaced coherent states, we have considered the state fidelity $F=\left|\langle\text{cat}(\lambda)|\psi'\rangle_m\right|^2$ between the post-measurement state of the mechanical mode and the cat state $\ket{\mathrm{cat}(\lambda)} \propto \ket{\lambda} + \ket{-\lambda}$ with $\ket{\pm\lambda}$ two coherent states of opposite phase and amplitude $\lambda\in\mathbb{C}$ that we treat as an unknown parameter. Maximizing $F$ with respect to the coherent state amplitude $\lambda$ of the cat state for $x=\bar{x}$ and the values of the other parameters used in Fig.~\ref{fig:hdyneWigner} {\bf (a)}, we find a maximum of 0.592 for $\operatorname{Re}(\lambda) \simeq0$ and $\operatorname{Im}(\lambda) = 0.715$. Such fidelity decreases to 0.5649 and 0.5646 for the parameters corresponding to the Wigner functions in Fig.~\ref{fig:hdyneWigner} {\bf (b)} and {\bf (c)} respectively. 

\begin{figure}[t!]
  {\bf (a)}
    \includegraphics[width=\columnwidth]{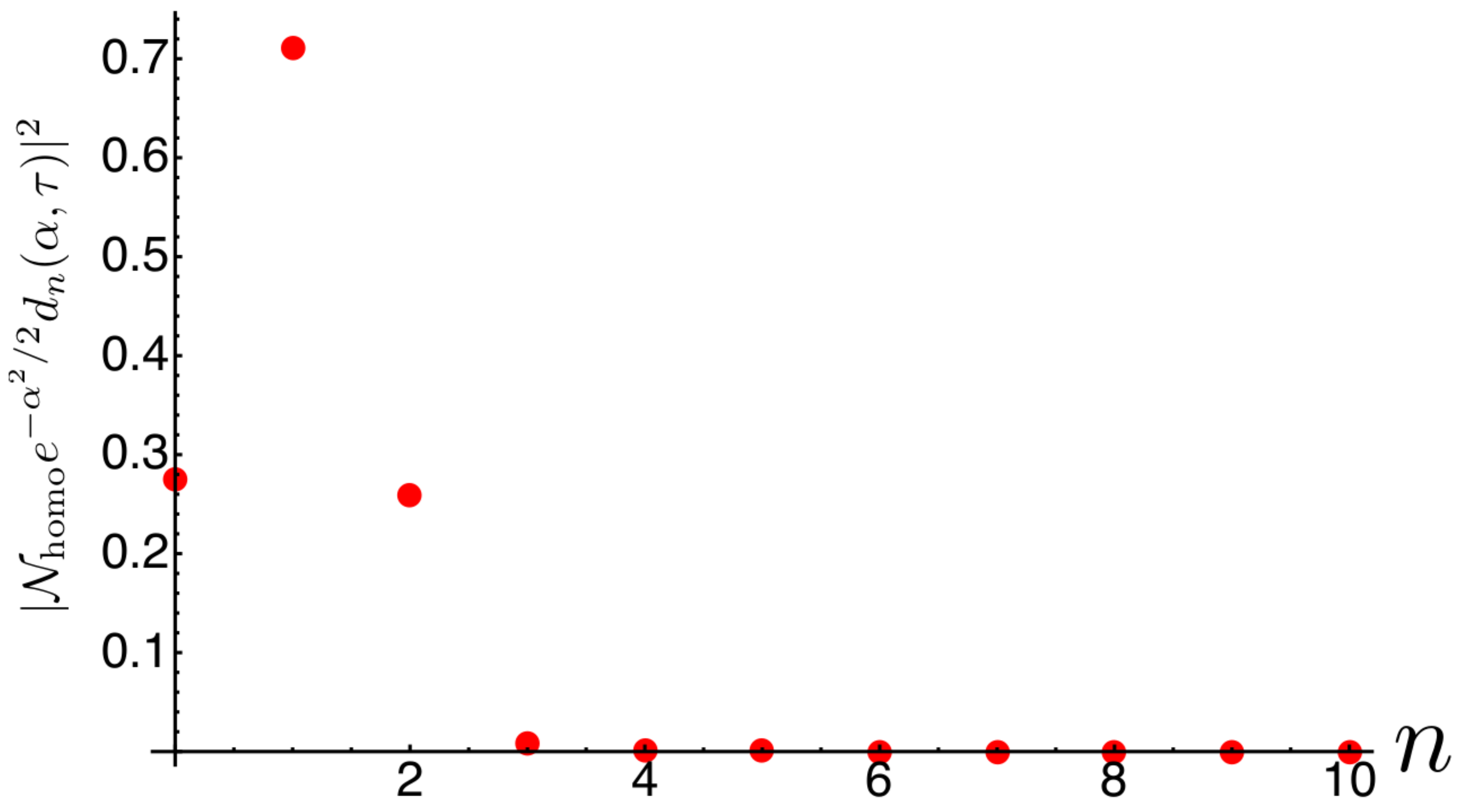}\\
   {\bf (b)}
    \includegraphics[width=\columnwidth]{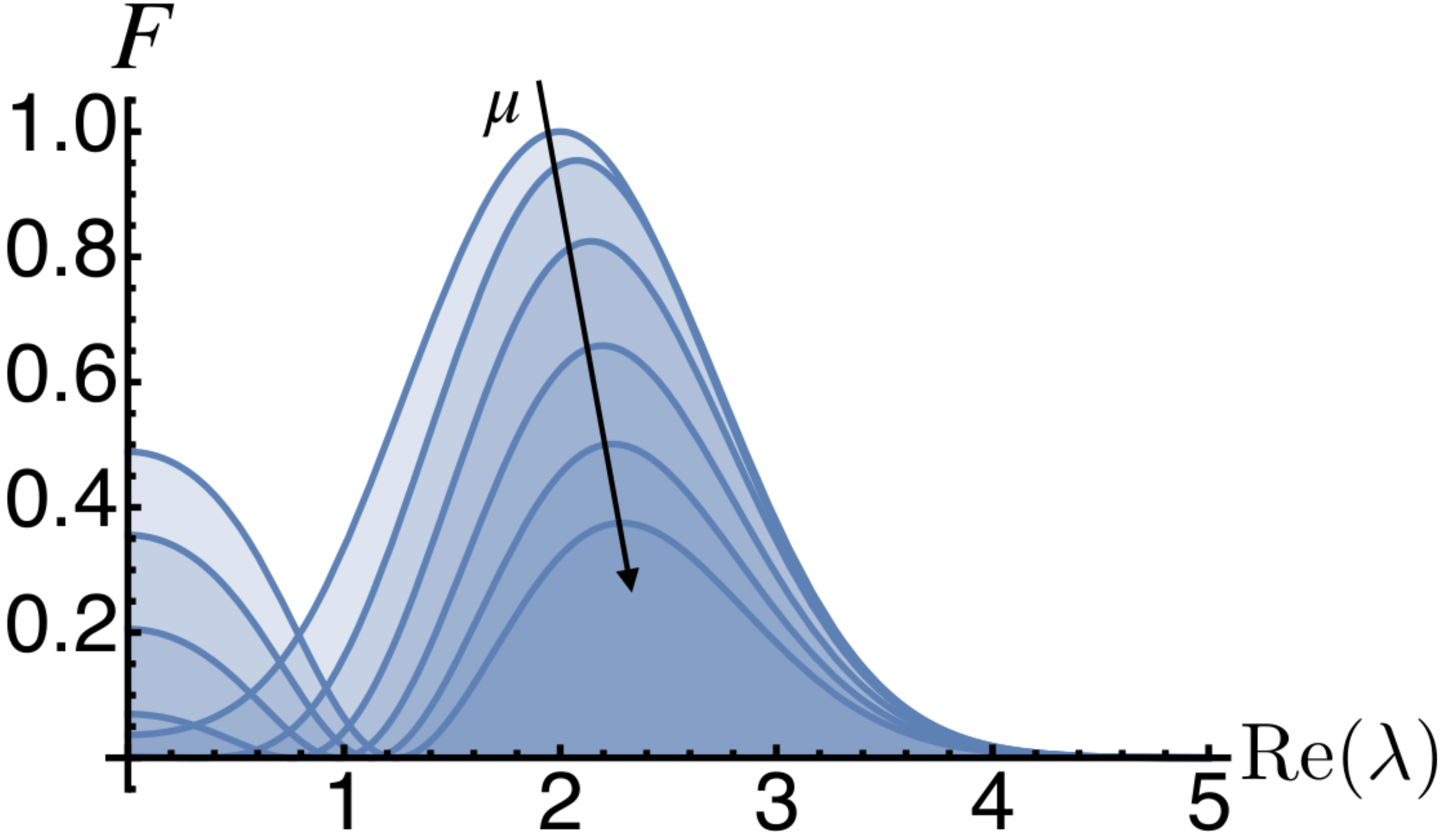}
    \caption{{\bf (a)}: Plot of the coefficients $|{\cal N}_\text{hom} e^{-\alpha^2/2}d_n(\alpha,\tau)|^2$ against $n$ for a mechanical state conditioned by homodyning the cavity field and post-selecting the outcome $x=\bar{x}$. We have taken $\tau=\pi$, $k=1$, $\beta=2$, $\theta=0$, and $\alpha=0.8$. {\bf (b)}: State fidelity $F$ between the conditional state of the mechanical mode considered in panel {\bf (a)} and state $\ket{C}$ for $\text{Im}[\lambda]=0$ and against $\text{Re}[\lambda]$ for $\mu=0,..,1$ varying in steps of 0.2 as indicated by black arrow. }
    \label{fig:coeffsq}
\end{figure}

While this result gives evidence of the decreased {\it cat-like} nature of states with non-zero $\theta$, the value of $F$ achieved for $\theta=0$ is somehow disappointing. However, a simple analysis allows us to interpret such result. We shall consider a simplified version of the conditional state of the mechanical mode consisting only of the most relevant components in Eq.~\eqref{general}. These are easily identified by looking at the  coefficients $|{\cal N}_\text{hom} e^{-\alpha^2/2}d_n(\alpha,\tau)|^2$. Fig.~\ref{fig:coeffsq} {\bf (a)} shows them for $n=0,\dots,10$, evidencing that all coefficients associated with $n \ge 3$ are in fact negligibly small. We can therefore consider state $\ket{C}$ made up of only three terms. Moreover, the value of the coefficient for $n=1$ is much higher than the that of $n=0$ and 2, which are roughly the same. We thus consider the simplified conditional state $\ket{C}_m$ given by
\begin{equation}
\begin{aligned}
    \ket{C}_m &= {\cal N'}_\text{hom} \left( d_0 (\alpha,\tau) \ket{\phi_0 (\tau)}_m + d_2 (\alpha,\tau) \ket{\phi_2 (\tau)}_m \right.\\
    &\left.+ \mu d_1 (\alpha,\tau) \ket{\phi_1 (\tau)}_m \right)
\end{aligned}
\end{equation}
with ${\cal N}'_\text{hom}$ the normalization constant and $\mu$ a weight parameter that we add to gauge the effect that $\ket{\phi_1(\tau)}_m$ has on the fidelity with a cat state. In particular, we look for the value that $\mu$ and $\lambda$ should take in order for the fidelity to be maximum. Our numerical analysis shows that one should take $\text{Im}[\lambda]=0$ and reduce $\mu$ as much as possible in order to have a state fidelity close to $1$ [cf. Fig.~\ref{fig:coeffsq} {\bf (b)}]. It is thus clear that the non-ideal fidelity highlighted above is the result of the contribution given by $\ket{\phi_1(\tau)}_m$ to the conditional state of the mechanical mode. 

\subsection{Effect of Heterodyne Detection} \label{hetero}

We have performed a similar analysis when, this time, the cavity field is subjected to ideal heterodyne measurements. As mentioned in Sec.~\ref{sec:hdetMirror}, these can be formally described as von Neumann projections onto coherent states of arbitrary amplitude $\sigma\in\mathbb{C}$, which deliver a conditional state of the mechanical mode reading
\begin{equation}
\ket{\psi'}_m={\cal N}_\text{het} e^{-|\alpha|^2/2}\sum^\infty_{n=0}c_n(\alpha,\tau)f_\text{het}(n)\ket{\phi_n(\tau)}_m
\end{equation}
with $f_\text{het}(n)$ as given in Eq.~\eqref{fM}. Scope of the analysis reported here is not only to infer the effectiveness of this conditional scheme to generate macroscopic quantum superpositions of the state of the mechanical mode, but also to evaluate the relative performance with the scheme based on homodyning. 

\begin{figure}[t!]
    \includegraphics[width=\columnwidth]{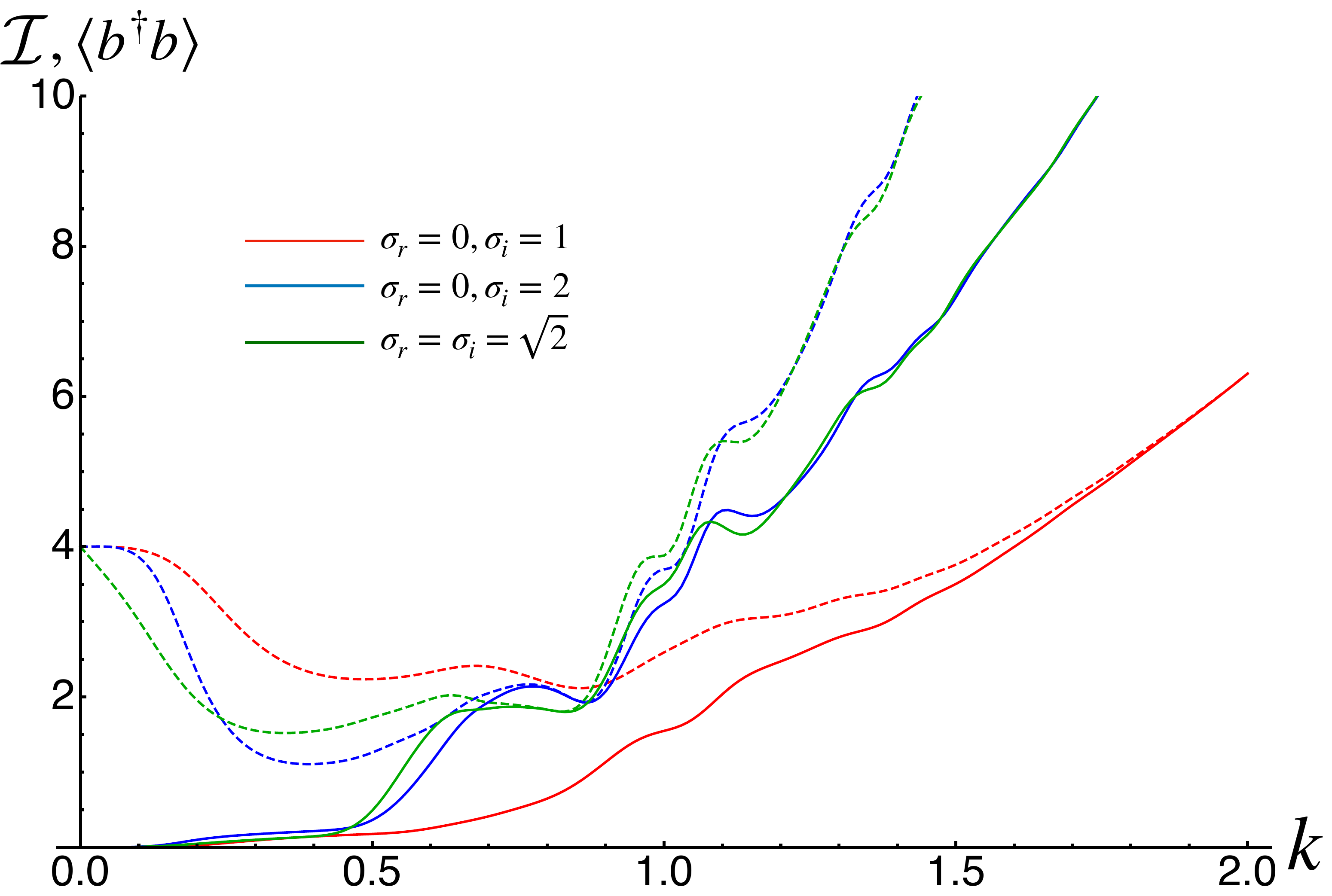}
    \caption{We plot the measure of macroscopicity ${\cal I}$ (solid lines) and the mean number of excitations (dashed lines) against $k$ and for various choices of the post-selected outcome $\sigma=\sigma_r+i\sigma_i$ of an ideal heterodyne measurement over the cavity field. The legend shows the values of $\sigma$ chosen in our simulations. All other parameters as in previous Figures.}
    \label{fig:hetero}
\end{figure}

\begin{figure*}[t!]
    \centering
    {\bf (a)}\hskip5.5cm{\bf (b)}\hskip5.5cm{\bf (c)}
    \includegraphics[width=2\columnwidth]{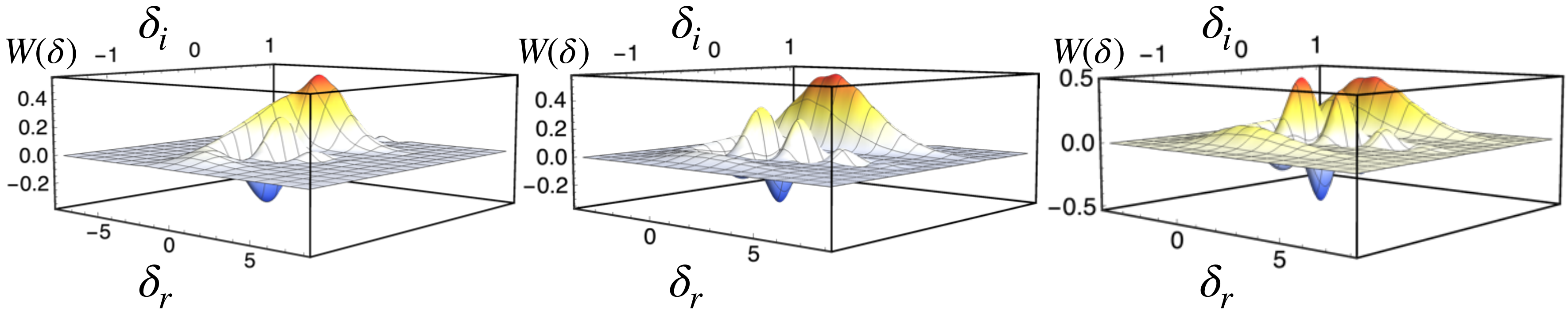}
    \caption{We report the Wigner function associated with the state of the mechanical system for the three cases addressed in Fig.~\ref{fig:hetero} and $k=1$.}
    \label{fig:heterodyneWigner}
\end{figure*}

In Fig.~\ref{fig:hetero} we report a summary of the results that we have gathered by considering increasing values of the effective coupling strength $k$ and various choices of the amplitude $\sigma=\sigma_r+i\sigma_i$. A first consideration to make is as follows: when looking at the same range of $k$ as in Fig.~\ref{fig:hdyneTheta} {\bf (a)}, the value of ${\cal I}$ achieved for a homodyne measurement with $x=\theta=0$ is comparable with what is found for a heterodyne measurement with $\sigma_r=0,\,\sigma_i=1$. We can thus say that projections onto $\ket{x(0)=0}$ and $\ket{\sigma=i}$ embody comparable resources, as far as the degree of macroscopic quantumness is concerned. However, contrary to the homodyne-measurement case -- where the gap between degree of macroscopicity and mean number of excitations remains wide for any $k\le{2}$, heterodyning returns states such that ${\cal I}\simeq\langle b^\dag b\rangle$. In this sense, one can legitimately claim for the superiority of a strategy based on heterodyning.

Exploring further the phenomenology resulting from heterodyne measurements, one intuitively expects that larger values of $|\sigma|$ could lead to larger degrees of macroscopicity ${\cal I}$ in light of the linear dependence of Eq.~\eqref{fM} from $\sigma^*$. This appears to be the case, as highlighted in Fig.~\ref{fig:hetero}, where it is also shown that the degree of macroscopicity is strongly determined by the choice of $|\sigma|$ and not of its real and imaginary parts per se.  

The calculation of the Wigner function associated with the state of the conditional mechanical oscillator for the three cases addressed in Fig.~\ref{fig:hetero} leads to the distributions reported in Fig.~\ref{fig:heterodyneWigner}, which are endowed with significant quantum coherences, as witnessed by the pronounced interference fringes in phase space. The different shapes taken by the Wigner functions as $\sigma$ changes can be justified through the analysis of the distribution of coefficients $|{\cal N}_\text{het} e^{-\alpha^2/2}d_n(\alpha,\tau)|^2$ against $n$: small values of $|\sigma|$ give rise to coefficients that are monotonically decreasing with $n$, so that the superposition in Eq.~\eqref{general} is strongly biased towards $\ket{\phi_0(\alpha,\tau)}$. As $|\sigma|$ grows, a structure similar to the one shown in Fig.~\ref{fig:coeffsq} {\bf (a)} emerges, leading to an increasing cat-like nature of the state of the mechanical mode. An analysis similar to the one performed before for homodyne measurements shows that for $|\sigma|=1.75$ the state of the mechanical oscillator can be very well approximated by the three-component superposition 
\begin{equation}
\ket{\psi'}_m\simeq{\cal N}_\text{het}e^{-|\alpha|^2/2}\sum^2_{j=0}\left(d_j(\alpha,\tau)\ket{\phi_j(\tau)}_m\right)
\end{equation}
with $d_1(\tau)$ being purely imaginary and $d_0(\tau),\,d_2(\tau)\in\mathbb{R}$ such that $|d_1(\tau)|\gg d_0(\tau)\simeq -d_2(\tau)$. The corresponding fidelity with a cat state of the form $\ket{\lambda}-\ket{-\lambda}$ is as large as $99.92\%$ for $\lambda\simeq2.003$.

\section{Effect of cavity dissipation on the degree of macroscopicity} 
\label{sec:noiselin}

We now address the effect that losses affecting the cavity have on the degree of macroscopicity of the mechanical mode, thus deviating from the unitary picture discussed so far. We shall restrict our attention to cavity dissipation, without worry about mechanical damping, as the latter typically occurs in timescales that are much longer than the decay of the cavity field owing to the large mechanical quality factors that can be achieved in current state-of-the-art optomechanical experiments. We thus model the open-system dynamics of the overall system through the master equation 
\begin{equation}
\label{ME}
    \frac{d \rho}{dt} = - \frac{i}{\hbar} [H, \rho] + \frac{\kappa}{2}(2a\rho a^\dagger- \{a^\dagger a, \rho\}) 
\end{equation}
where $\kappa$ is the decay rate of the cavity field and $H$ is the Hamiltonian in Eq.~\eqref{HamLinear} in the interaction picture. 

Guided by the analysis performed in the previous Sections, we shall retain a value of the effective coupling strength of the order of $k\sim1$. This entails that, in order to make our calculations accurate, we should retain values up to $n=10$ in Eq.~\eqref{timeEvState}, thus making the direct solution of Eq.~\eqref{ME} a difficult problem. We thus resort to a quantum unravelling approach~\cite{properQU} based on the chopping of the time window within which we are interested in tracking the dynamics of the system in $n\gg1$ small time steps $\delta\tau$.  At the $n^\text{th}$ time step (such that $n\delta\tau=\tau$), we calculate the probability that a quantum jump of the state $\ket{\psi(\tau)}$ occurs, which is evaluated as 
\begin{equation}
    \delta p = \kappa \delta\tau \bra{\psi(\tau)} a^\dagger a \ket{\psi(\tau)}.
\end{equation}
Such probability is typically much less than 1. We then generate a random number $\epsilon\in[0,1]$: should we have $\epsilon>\delta p$, the state of the system at time $\tau + \delta \tau$ becomes
\begin{equation}
    \ket{\psi(\tau+\delta\tau)} = (\one - i \tilde{H} \delta \tau)\ket{\psi(\tau)},
\end{equation}
where $\tilde{H}$ is the non-Hermitian interaction-picture Hamiltonian
\begin{equation}
    \tilde{H} = - k a^\dagger a ( b e^{i \tau} + b^\dagger e^{-i \tau} ) - i \kappa a^\dagger a.
\end{equation}
In the less probable case of $\delta p> \epsilon$t, a quantum jump described by the operator $C = \sqrt{\kappa} a$ occurs, effectively leaking a photon of the cavity. This algorithm is repeated at every time step to achieve a dynamical trajectory. Many such trajectories are then constructed and the state of the system built as their ensemble average to remove any dependence over the realization of the sequence of quantum jumps. 

\begin{figure}[t!]
    \centering
    \includegraphics[width=\columnwidth]{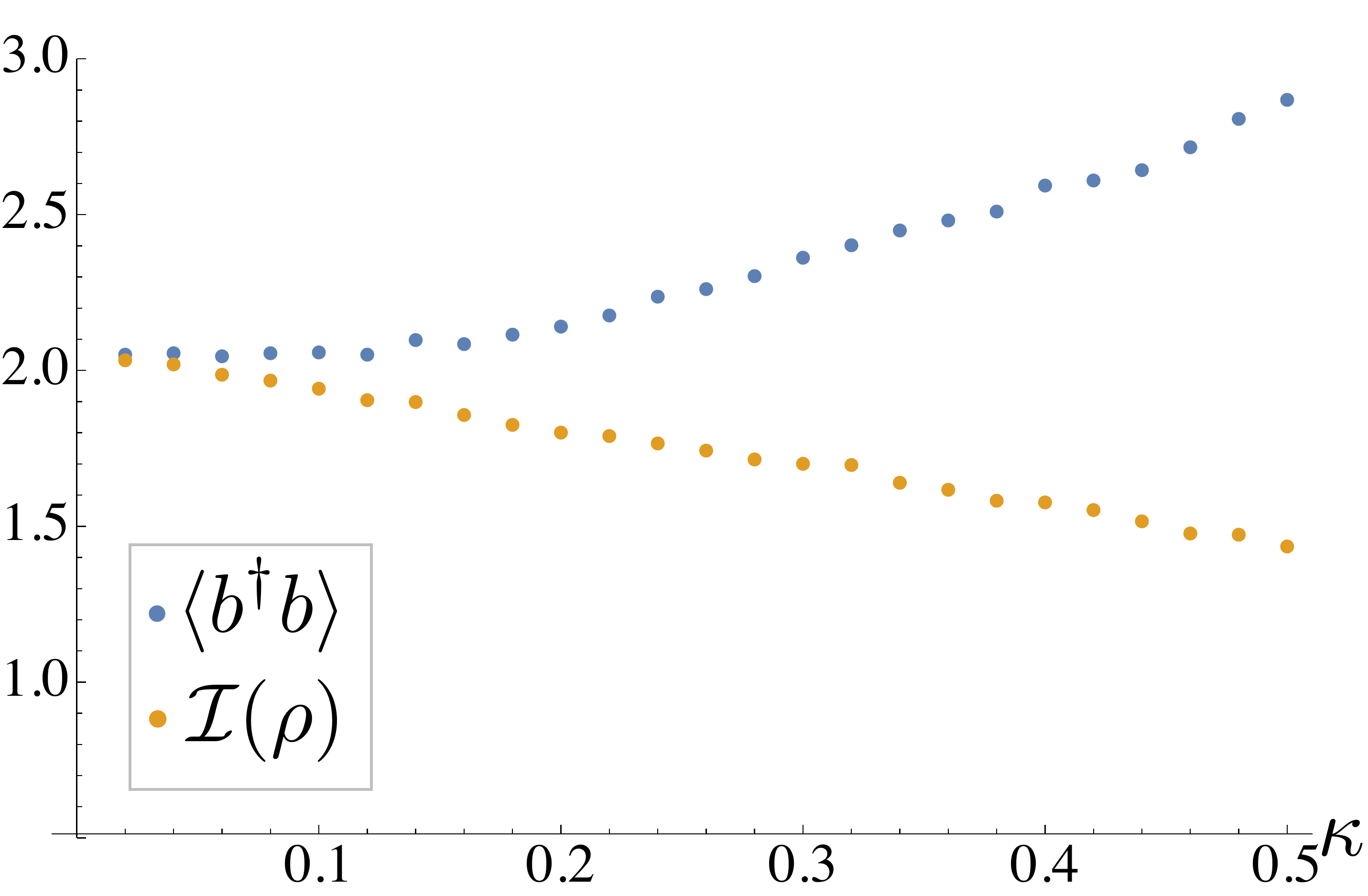}
    \caption{As the cavity damping rate $\kappa$ increases (in units of $\omega_m$), the degree of macroscopic quantumness decreases and moves further away from its upper bound embodied by the mean number of mechanical excitations. In this plot we have used the parameters $\alpha=0.8$, $\beta=2$, $k=1$, $\tau=\pi$ and $\delta \tau=\pi \times 10^{-5}$.}
    \label{fig:quPlotLin}
\end{figure}

The unraveling strategy is used to address the effect that a growing cavity decay rate has on the macroscopic nature of the conditional mechanical state achieved when performing homodyne measurements (no qualitative difference is found when heterodyne measurements are chosen). The results are reported in Fig.~\ref{fig:quPlotLin}, where $\kappa$ is measured in units of $\omega_m$. The cavity dissipation evidently induces a decrease of the degree of macroscopicity achieved for $\kappa=0$, $\theta=0$ and the same parameters used in the simulations reported in Fig.~\ref{fig:hdyneTheta}. Moreover, we find a bifurcation between the mean number of excitations in the mechanical state and ${\cal I}(\rho)$: as the cavity damping rate increases, the state of the mechanical system becomes more mixed than in the unitary-dynamics case, thus magnifying the difference between the actual degree of macroscopicity and its maximum achievable value.

How does this result depend on the initial state of the mechanical system? In order to address this question we generalized the assumptions made in the previous Sections and consider the mechanical mode as initially prepared in the displaced thermal state with an amplitude of displacement equal to the parameter $\beta$ used earlier. 

\begin{figure}[t!]
    \centering
    \includegraphics[width=\columnwidth]{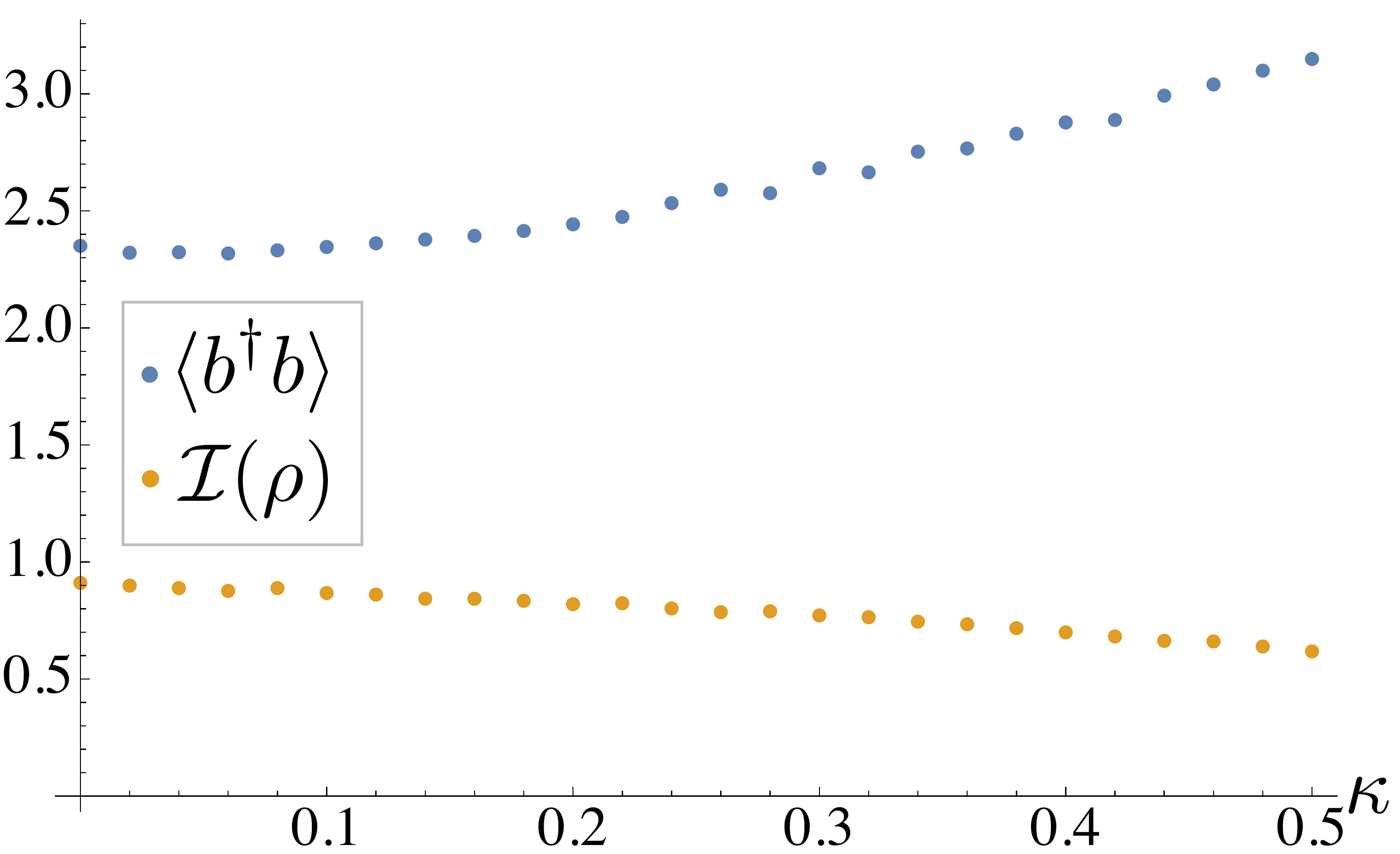}
    \caption{Macroscopic quantumness (yellow dots) and $\langle b^\dagger b\rangle$ (blue dots) for a mechanical mode initially in a  displaced thermal state. The higher the cavity damping rate $\kappa$ (in units of $\omega_m$), the lower the macroscopic quantumness of the state and the greater the difference between the macroscopicity and its theoretical upper bound.}
    \label{fig:thermal}
\end{figure}

The degree of macroscopicity of the state of the mechanical mode after homodyne detection is plotted against the cavity damping rate in Fig.~\ref{fig:thermal}. Indeed, the macroscopic quantumness of the state is much less than when the mirror was initially in a coherent state. In this case, there is never a point where the macroscopic quantumness is close to the upper bound provided by the mean number of excitations in the mechanical state. 

\begin{figure}[t]
{\bf (a)}\hskip4cm{\bf (b)}\\
        \includegraphics[width=0.49\columnwidth]{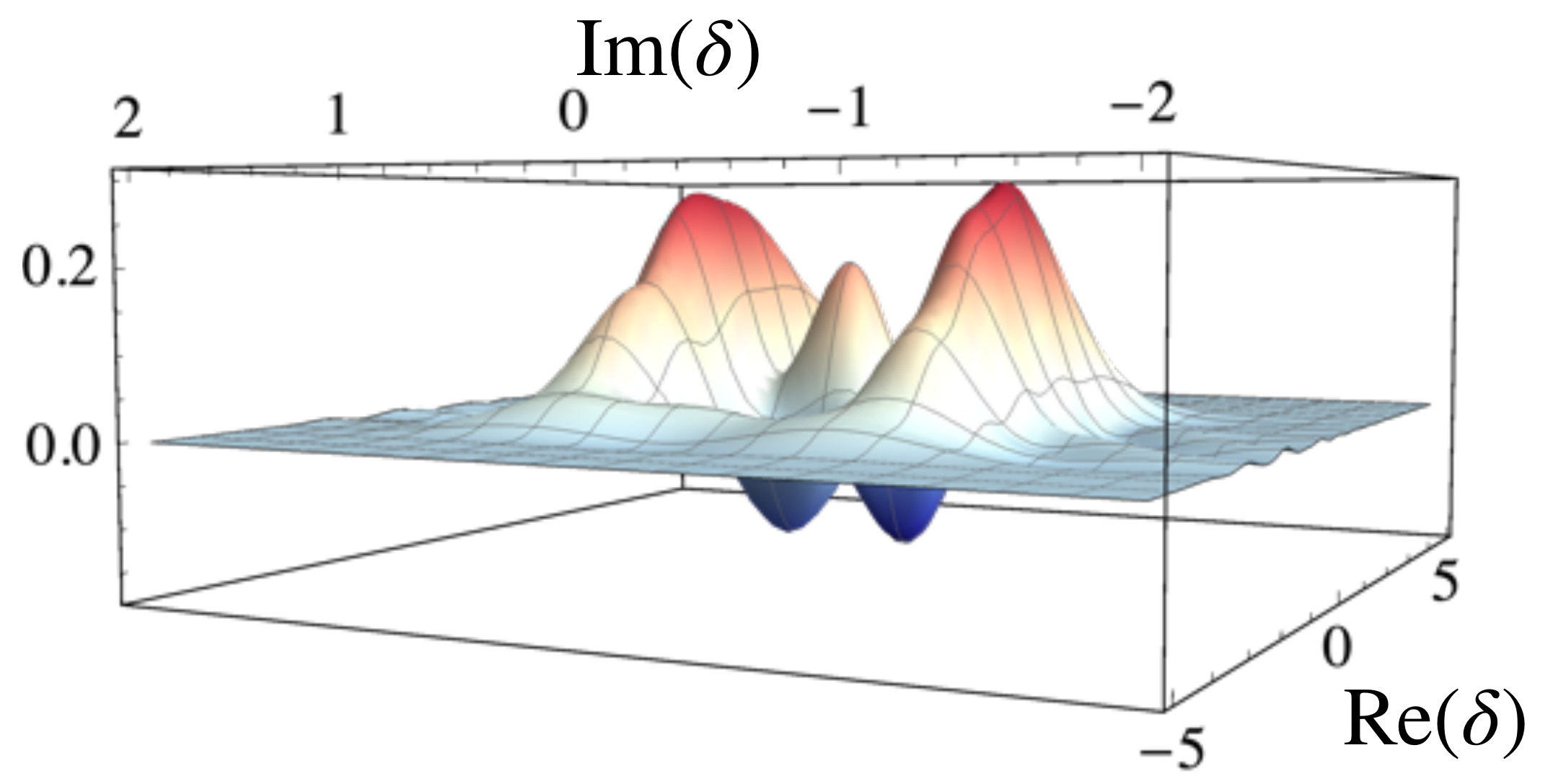}
        \label{fig:mean and std of net14}
        \includegraphics[width=0.49\columnwidth]{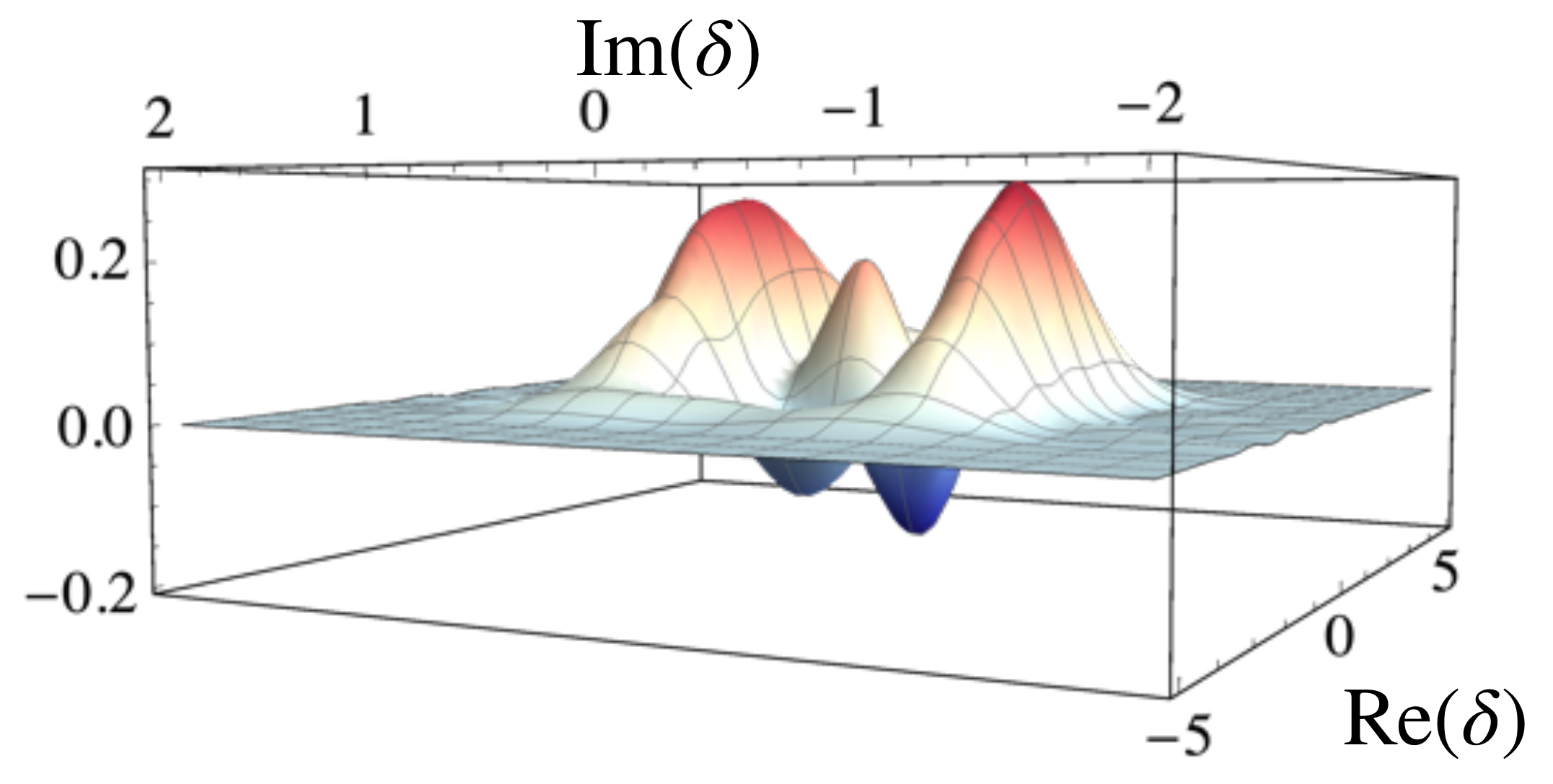}
        \label{fig:mean and std of net24}
        {\bf (c)}\hskip4cm{\bf (d)}\\
        \includegraphics[width=0.49\columnwidth]{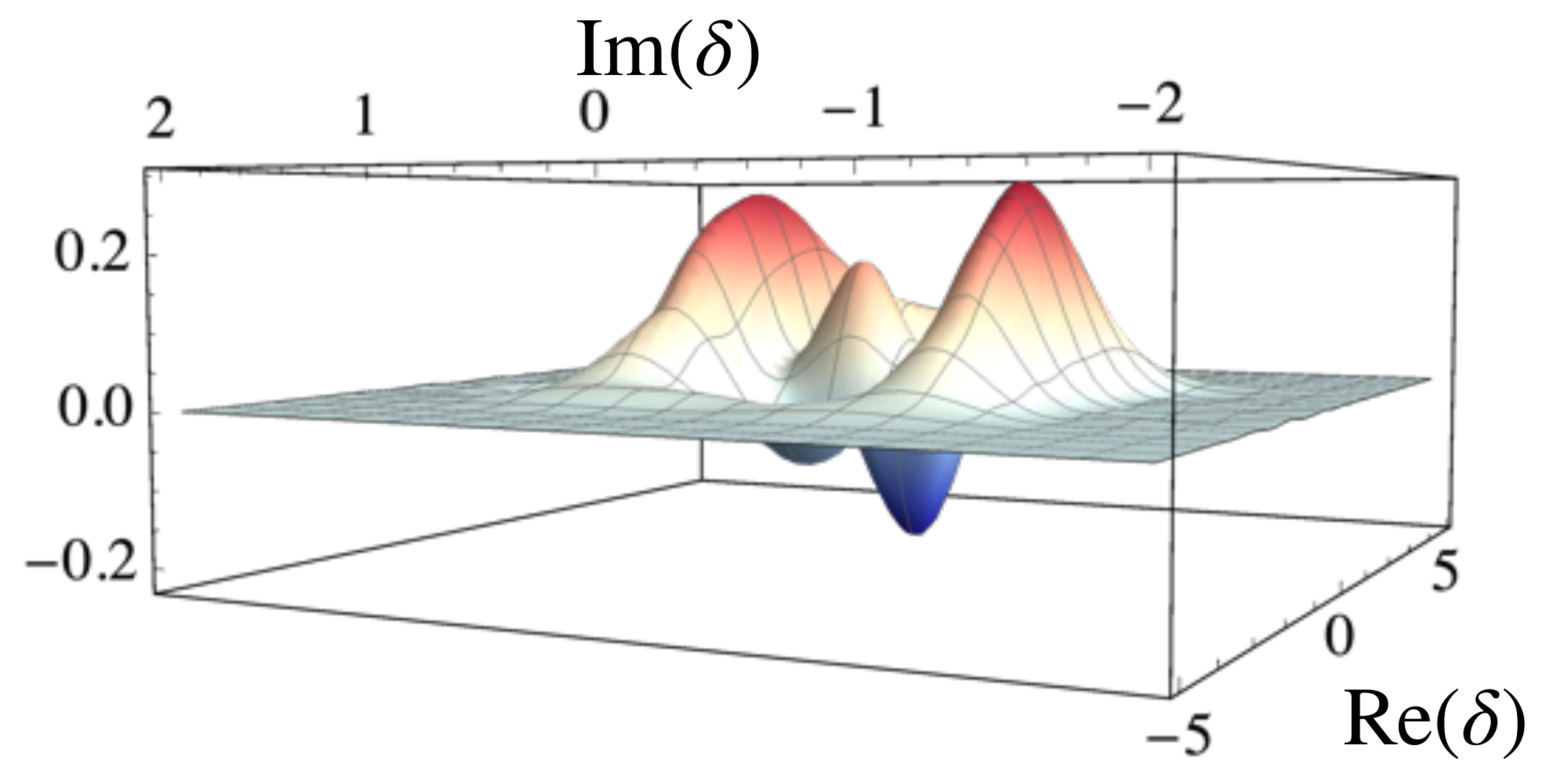}
        \label{fig:mean and std of net34}
        \includegraphics[width=0.49\columnwidth]{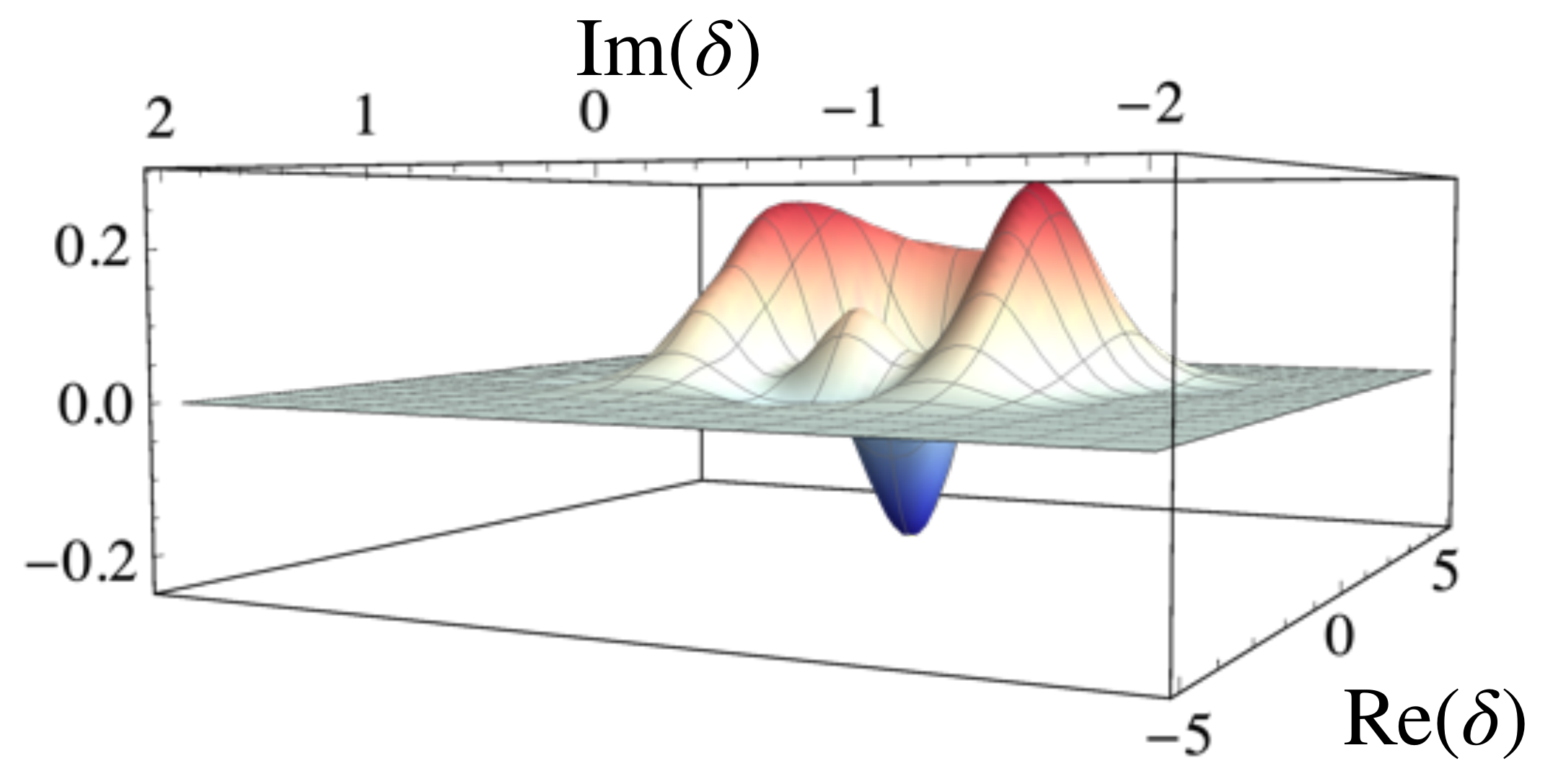}
        \label{fig:mean and std of net44}
    \caption{Wigner functions of the time evolved state of the mirror from an initial thermal state when $\kappa/\omega_m = 0.02, 0.1, 0.2$ and 0.5 respectively.} 
    \label{fig:thermalWignerFns}
\end{figure}

Instances of the shape taken by the Wigner function of the mechanical system as $\kappa$ increases are given in Fig.~\ref{fig:thermalWignerFns}, which illustrates well the causes of the decreased value of ${\cal I}$: the Gaussian peaks representing the distinguishable components in the cat-like state engineered by our conditional approach are made to merge by the increasing damping mechanism, while the interference fringes become less prominent. We can still see negativity in the Wigner function for $\kappa/\omega_m$ up to $0.5$, thus demonstrating a residual quantum nature of the corresponding state. However both the amplitude and the frequency of the fringes are depleted, thus resulting in lower values of ${\cal I}$. 

\section{Macroscopicity of excitation-subtracted mechanical states} \label{sec:subtraction}

In this Section we address another class of non-classical states achieved by conditioning the mechanical part of an optomechanical system by means of measurements performed on the cavity field. Specifically, we consider the reduced mechanical states achieved by subtracting one photon from the field, thus following the scheme proposed in Ref.~\cite{MauroPRL} and developed further in~\cite{Miskeen}. Such a scheme is able to generate non-classical states of the mechanical mode that, under suitable operating conditions, exhibit a pronounced cat-like form while resembling a single-excitation number state at low temperature and for a low-mass mechanical mode.

While we refer to Ref.~\cite{MauroPRL,Miskeen} for details on the protocol itself, it is sufficient to mention here that the conditional state $\rho_{ps}$ of the mechanical system achieved upon subjecting the cavity field to the subtraction of a single photon can be written as
\begin{equation}
\label{rhops}
\rho_{ps}=\frac{{\cal N}}{\pi^2}\int\, \chi'(\gamma)D(-\gamma)d^2\gamma
\end{equation}
with ${\cal N}$ a normalization constant and $\chi'(\gamma)$ the characteristic function of the reduced state of the mechanical system, conditioned on the subtraction of a single quantum from the state of the field. Its form is immaterial for the scopes of this manuscript and can be retrieved from Refs.~\cite{MauroPRL,Miskeen}. The Wigner function corresponding to such state is readily cast in the form
\begin{equation}
W_{ps}(\delta)={\cal N} {\cal A}(\delta)e^{-2 {\bm \delta}{\bm \sigma}^{-1}{\bm \delta}^T}
\end{equation}
with ${\bm\delta}=(\text{Re}(\delta),\,\text{Im}(\delta))$ the vector of phase-space variables, ${\bm\sigma}=\text{diag}[(\Delta x)^2,(\Delta p)^2]$ the matrix of variances of the quadrature operators of the mechanical mode, and ${\cal A}(\delta)$ a second-degree polynomial of variable $\delta$. Depending on the parameters of the problem, ${\cal A}(\delta)$ can take negative values in certain regions of the phase space, thus leading to the possibility to enforce mechanical non-classicality through photon subtraction. 

\begin{figure}
\includegraphics[width=\columnwidth]{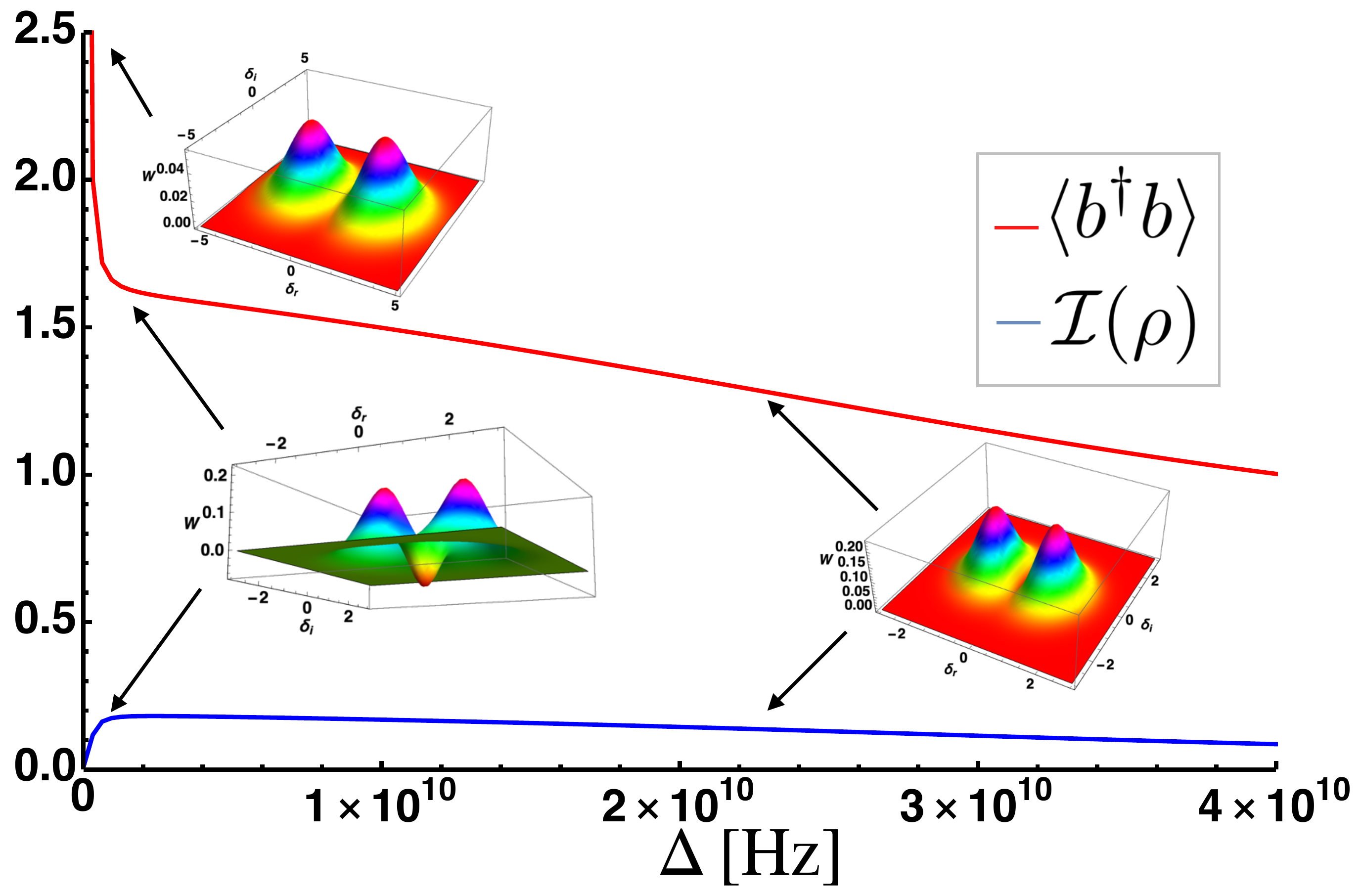}
\caption{Measure of macroscopicity $\mathcal{I}$ (blue) and $\langle b^\dagger b\rangle$ (red) as detuning between the external pump and cavity field $\Delta$ increases. The Wigner function of the mechanical state is shown for various values of $\Delta$ including the point where $\mathcal{I}$ is maximum. At this point we can see negativity in the Wigner function indicating the quantum nature of the state.}
\label{MacroSubtract}
\end{figure}

The evaluation of the measure of macroscopic quantumness is then performed using the alternative expression for ${\cal I}(\rho)$ given in Ref.~\cite{Macro}
\begin{equation}
{\cal I}(\rho_{ps})=-\frac{\pi}{2}\int\,d^2\delta\,W_{ps}(\delta){\partial^2_{\delta\delta^*}}W_{ps}(\delta),
\end{equation}
while the evaluation of the mean number of excitations in the state of the mechanical mode makes use of the expression 
\begin{equation}
\label{meanWig}
\langle b^\dag b\rangle=\int\,|\delta|^2 W_{ps}(\delta)\,d^2\delta-\frac{1}{2},
\end{equation}
which can be easily proven following Ref.~\cite{Alessandro} and the lines sketched in the Appendix. The results of our analysis are reported in Fig.~\ref{MacroSubtract}, where we study the measure of macroscopicity against the detuning $\Delta$ between external pump and cavity field in relation to the Wigner function of the mechanical mode. There is clearly a working point -- corresponding to small values of $\Delta$ with respect to the cavity frequency -- at which the mechanical mode achieves the largest macroscopic quantumness allowed by the conditions of its dynamics, although such value is far from being the maximum possible one, as it can be appreciated by comparing ${\cal I}(\rho_{ps})$ to the red curve in Fig.~\ref{MacroSubtract}, which shows the mean number of excitations in the reduced mechanical state. Such a discrepancy is easily justified by the fact that the initial state of the mechanical system considered in this analysis is mixed.

\section{Conclusions} \label{sec:conc}

We have studied the relative performance of conditional measurements that can be made on a cavity field in an optomechanical system to engineer macroscopic quantum states of a mechanical mode. We focused on the influence that the measurement settings of general homodyne and heterodyne measurements performed on the light field of an optomechanical system have on the degree of macroscopicity of the mechanical state. Both classes of Gaussian measurements are successful in generating macroscopic quantum states, steering the mechanical mode towards states that strongly resemble Schr\"{o}dinger cat states. However, qualitatively significant differences emerge from different choices of general-dyne measurements performed on the light field, showing that the specific conditional choice to apply in order to generate a macroscopically quantum state should be gauged against the operating conditions of the optomechanical system. Furthermore, the quantum nature of the state persists even in the presence of dissipative processes, making the conditional scheme for mechanical macroscopicity robust. Secondly, we studied the effect of a scheme based on photon subtraction on the degree of macroscopic quantumness of the mechanical mode finding working points at which it is possible to generate a mechanical state exhibiting macroscopic quantum features, although the degree of macroscopic quantumness remains comparatively low. Our study, which is based on a general optomechanical system [cf. Eq.~(\ref{HamLinear})], provides results that are relevant for a range of different experimental settings~\cite{AspelmeyerReview}, including cavity-less platforms where the ability to reconstruct quasi-probability functions has been demonstrated~\cite{Rashid2017}.

\acknowledgments 
Support by the H2020 Collaborative Project TEQ (Grant Agreement 766900), the SFI-DfE Investigator Programme (grant 15/IA/2864), the Leverhulme Trust Research Project Grant UltraQuTe (grant nr. RGP-2018-266), and the Royal Society International Exchanges Programme (IEC\textbackslash R2\textbackslash192220) is gratefully acknowledged. MP is also supported by a Royal Society Wolfson Fellowship (RSWF\textbackslash R3\textbackslash183013).

\section*{Appendix}

Here we provide details on the calculation of the expression in Eq.~\eqref{meanWig}, which can be deduced from the calculation of the Wigner function of operator $b^\dag b$ and the use of the trace rule in phase space. We start from Eq.~\eqref{rhops}, from which we have
\begin{equation}
\label{mean}
\langle{b}^\dag b\rangle=\frac{{\cal N}}{\pi^2}\int\, \chi'(\gamma)\text{Tr}[b^\dag bD(-\gamma)]d^2\gamma.
\end{equation}
The trace within the integral can be evaluated using the cyclic property of trace, the commutation relation $[b,b^\dag]=\openone$ and the over-complete basis of coherent states as follows
\begin{equation}
\begin{aligned}
\text{Tr}[b^\dag bD(-\gamma)]&\equiv\text{Tr}[b^\dag D(-\gamma)b]-\text{Tr}[D(-\gamma)]\\
&=\text{Tr}[b^\dag D(-\gamma)b]-\pi\delta^2(-\gamma)\\
&=\frac{1}{\pi}\int\,|\beta|^2e^{-|\gamma|^2/2+\gamma^*\beta-\gamma\beta^*}\,d^2\beta-\pi\delta^2(-\gamma).
\end{aligned}
\end{equation}
When used in Eq.~\eqref{mean} with $\chi'(\gamma)$ replaced by the anti-Fourier transform of the Wigner function $W_{ps}(\delta)$, one find the expression in Eq.~\eqref{meanWig}.

\bibliographystyle{apsrev4-1}
\bibliography{macroPaperRefs-1.bib}

\end{document}